\documentclass[twocolumn,aps,prc,superscriptaddress,showpacs]{revtex4}
\usepackage{amssymb}
\usepackage{amsmath,bm}
\usepackage{graphicx}
\usepackage{dcolumn}
\usepackage{multirow}
\usepackage{color}
\usepackage{CJK}
\usepackage{cases}
\usepackage{subfigure}
\begin{document}
\begin{CJK*}{GBK}{song}
\draft
\title{Crossover transition in the Fluctuation of Internet}

\author{Jiang-Hai Qian}
\affiliation{\footnotesize Shanghai Institute of Applied Physics,
Chinese Academy of Sciences, Shanghai 201800, China}
\author{ Qu Chen}
\affiliation{\footnotesize School of Information Science and Technology, East China
Normal University, Shanghai 200241, China}
\author{ Ding-Ding Han}
\affiliation{\footnotesize School of Information Science and
Technology, East China Normal University, Shanghai 200241, China}

\author{ Yu-Gang Ma}


\affiliation{\footnotesize Shanghai Institute of Applied Physics,
Chinese Academy of Sciences,  Shanghai 201800, China}


\date{\today}
\nopagebreak

\begin{abstract}

Gibrat's law predicts that the standard deviation of the growth rate
of a node's degree is constant. On the other hand, the preferential
attachment(PA) indicates that such standard deviation decreases with
initial degree as a power law of exponent $-0.5$. While both models
have been applied to Internet modeling, this inconsistency requires
the verification of their validation. Therefore we empirically study
the fluctuation of Internet of three different time intervals(daily,
monthly and yearly). We find a crossover transition from PA model to
Gibrat's law, which has never been reported. Specifically Gibrat-law
starts from small degree region and extends gradually with the
increase of the observed period. We determine the validated periods
for both models and find that the correlation between internal links
has large contribution to the emergence of Gibrat law. These
findings indicate neither PA nor Gibrat law is applicable to the
actual Internet, which requires a more complete model theory.
\end{abstract}
 \maketitle

\section{Introduction}
The evolution of Internet has long been a heat subject since the
dawn of complex network theory, due to its rich data, wide
application and nontrivial
properties~\cite{internet0,internet1,internet2,internet3,internet4,internet5,internet6,internet7,internet8,internet9,internet10}.
While models based on either stochastic process or optimal strategy
are continually proposed, an urgent question to be addressed is that
which of them is really applicable or is validated to describe the
actual evolution of Internet~\cite{internet0,va}. This question
concerns not only our understanding on the process of internet
evolution, but also the possibility of our further goal of control
and prediction of this large-scale system.

Most popular models of Internet base on the mechanism of
preferential attachment(PA), which describes that the probability of
a node to capture links is proportional to its current degree. It is
considered to be essential for producing the power-law degree
distribution~\cite{origin1,origin2,origin3,origin4,confirm1,confirm2,confirm3}.
While some evidences suggest PA is unapplicable for route-level
Internet~\cite{internet0}, other empirical studies based on
mean-field approach support its validation for AS level (autonomous
system level) and other types of
networks~\cite{confirm4,confirm5,confirm6}.

Another approach to model Internet follows its statistical law
instead of the detailed descriptions as PA~\cite{flu}. The
representative case is Gibrat law which has been introduced as the
candidate of Internet model to characterize the dynamics of the
constant appearance and disappearance of links and
nodes~\cite{gibrat network,internet8}. The traditional Gibrat law
assumes that the growth rate of a variable such as population, the
number of messages sent by a person or the degree of a node has an
independent identically distributed(i.i.d) structure so that both
its mean and standard deviation are independent of the initial value
of the variable~\cite{gibrat}. Although this assumption seems
rejected by a variety of recent empirical
studies~\cite{stane1,stane2,stane3,stane4,stane5,stane6,human}, it
succeeds in reproducing the exact power-law exponent of the degree
distribution of Internet~\cite{internet8}.

While the validation of both model is still controversial, a more
serious problem is that there exists an inconsistency even between
themselves. As is indicated in Ref~\cite{human} and will be
specified in section II in the present paper, the conditional
standard deviation of degree growth rate of PA decays with initial
degree as a power law of exponent $-0.5$, which contradicts with
Gibrat assumption. This raises the question that which model is more
appropriate for describing the evolution of Internet not only at a
mean-field level but also on a fluctuation aspect. Unfortunately
previous empirical studies based on mean-field
method~\cite{confirm4,confirm5,confirm6} cannot distinguish PA and
Gibrat law since both them cause the similar proportionate effect.
While the fluctuation property may uncover some important nature of
Internet, it has been rarely empirically studied. The main purpose
of the present paper is to determine the actual fluctuation property
of Internet topology and the scope of the validation of the two
models, which is significant both theoretically and practically.

The paper is organized as follow. In section II we show the
inconsistency between PA and Gibrat law by deriving the relation
between the standard deviation of degree growth rate and initial
degree. In section III we empirically study the fluctuation of
Internet topology for three different time scale(daily, monthly and
yearly). We find that the fluctuation of internet experience a
crossover transition from PA model to Gibrat's law with the increase
of the observed period. We determine the validated period for both
PA and Gibrat's law respectively and discuss the possible cause of
the emergence of Gibrat law. In section IV we draw the conclusion.

\section{Inconsistency between Gibrat law and PA rule}
The proportionate effect described by Gibrat law can be formalized
by the following random multiplicative process:
\begin{equation}
k_i(t+1)=[1+\varepsilon_i(t)] k_i(t),
\end{equation}
where $k_i(t+1)$ and $k_i(t)$ are the degree of node $i$ at time
$t+1$ and $t$, and $\varepsilon_i(t)$ is a random process. The
degree growth rate is defined as
\begin{equation}
r_i=\log\frac{k_i(t+1)}{k_i(t)}.
\end{equation}
More generally, if we observe the system by interval $\Delta t$, the
growth rate $\log({{k_i(t+\Delta t)}/{k_i(t)}})$ is given by
\begin{equation}
r_i(\Delta t)\sim \sum_{j=t}^{t+\Delta t}\varepsilon_i(j) .
\end{equation}
The basic assumptions of Gibrat law are that $\varepsilon_i$ is
$(1)$ independent of its initial degree and $(2)$ uncorrelated in
time~\cite{gibrat}. The two assumptions indicate that the
fluctuation property of degree growth, characterized by the standard
deviation of $r_i(\Delta t)$ conditional to initial degree
$k_0=k_i(t)$ follows
\begin{equation}
\sigma_r(k_0)\sim const.
\end{equation}

On the other hand the fluctuation of degree growth of PA behaves
differently. The PA rule describes that the probability $p_k$ of a
new link to connect to a node relates to nothing else but the node's
current degree, which is given by
\begin{equation}
p_k\propto k.
\end{equation}
In other words the creation of links are uncorrelated with each
other and the evolution of degree is a memoryless Markov process. By
mean-field method, the evolution of the degree of a node is
$dk/dt=g(t)k$, where $g(t)$ is usually a function related to the
growth pattern of network size. Solving the equation we have
\begin{equation}
k(t)\propto \frac{G(t)}{G(\tau)},
\end{equation}
where $\tau$ is the birth time of the node and $G(t)=e^{\int
g(t)dt}$. Now Let us denote random variable $X(t)$ as the number of
new links connecting to a node at time $t$. Its i.i.d structure
indicates that it follows the Binomial distribution, whose variance
$\sigma^2_{X(t)}$ is proportional to $p_k(1-p_k)$. Considering
$p_k\ll1$, we have
\begin{equation}
\sigma^2_{X(t)}\sim p_k.
\end{equation}
The degree increment of the node from $t$ to $t+\Delta t$ is $\Delta
k=k(t+\Delta t)-k(t)=\sum_{i=t}^{t+\Delta t}X(i)$. According to the
definition of the growth rate $r$, we have
\begin{equation}
r(\Delta t)\sim \frac{\sum_{i=t}^{t+\Delta t}X(i)}{k(t)}.
\end{equation}
Reminding that the creation of links are uncorrelated in time, the
conditional variance of $r(\Delta t)$ is
\begin{equation}
\sigma_r^2(k(t))\sim \frac{1}{k(t)^2}\int_t^{t+\Delta
t}\sigma^2_{X(i)}di.
\end{equation}
Substituting Eq(5)$\thicksim$Eq(7) to Eq(9) and replacing $k(t)$
with $k_0$, we finally derive the fluctuation property of degree
growth rate for PA
\begin{equation}
\sigma_r(k_0)\sim k_0^{-0.5}.
\end{equation}
Note that Eq(10) is valid for other events such as rewiring and link
deletion as long as they do not break the memoryless property.

Eq(4) and Eq(10) indicate a basic contradiction between PA and
Gibrat law even though both of them are reported to be validated at
mean-field level. Our question is which model is closer to the
reality and what is the real fluctuation property of Internet on
earth. On the other hand PA and Gibrat law share common stationary
property in the sense that both the scaling properties of Eq(4) and
Eq(10) are independent of the observed period $\Delta t$. As will be
presented in the next section, neither of the models can totally
characterize the real fluctuation but is validated for two different
periods. With the increase of the periods, the fluctuation pattern
changes gradually, which contrasts to the stationary property of
both models.

\section{Empirical result and correlation analysis}
In this section we will empirically study the fluctuation of degree
growth rate of Internet and determine the periods $\Delta t$, for
which PA and Gibrat law are validated respectively. In addition we
will briefly discuss the origin of the emergence of Gibrat law.

Our empirical data come from the Oregon Route Views
project~\cite{data}. They include snapshots of three different time
scales, i.e. daily($30$ days: $2006/09/01\sim 2006/09/30$),
monthly($36$ months: $2005/01\sim 2007/12$) and yearly($15$ years:
$1998\sim 2012$). The original data are collected in the form of
Border Gateway Protocol routing tables, from which an Internet graph
can be constructed. As usual, each node represents a specific AS
while each edge is the logical link between the inter-connected
ASes, so that we obtain a network of size of O($10^5$) nodes and of
an almost constant average degree about $4.5$. The topological
properties that we measured are stationary for all the three time
scale and are consistent with previous empirical
studies~\cite{internet3,internet10}. The degree distribution is
power law as $p(k)\sim k^{-\alpha}$ with exponent $\alpha\approx 2.1
$. We also check the dynamics of the preferential attachment as done
in Ref~\cite{confirm4}. We find for all the three time scale, the
linear PA $\Delta k\sim k$ is always valid.

\begin{figure}[htbp]
\centering
\includegraphics[width=3.5in]{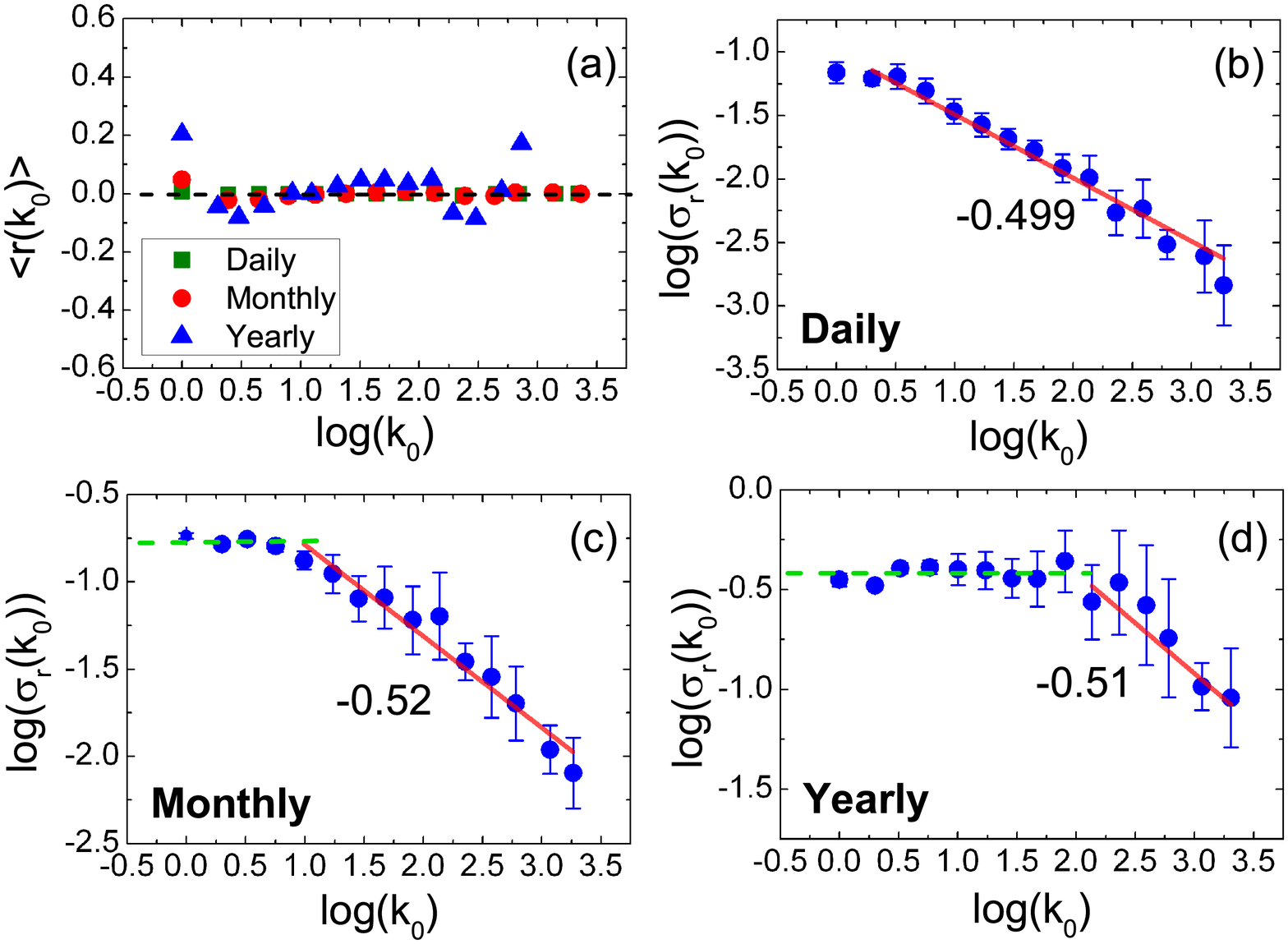}
\caption{Conditional average and conditional standard deviation of
the degree growth rate versus initial degree: (a)the conditional
average for three periods. All of them are independent of initial
degree $k_0$ and stay around constant $0$. (b)the conditional
standard deviation for daily data. It decreases with $k_0$ as power
law of exponents $-0.5$, as predicted by PA. (c)the conditional
deviation for monthly data. The small-degree region of $k_0<10$
becomes flat compared to daily data, while the rest of region
remains unchanged. (d)the conditional deviation for yearly data.
Gibrat law dominates for a large range of $k_0<300$. All the data
are logarithmic binned and are plotted on a log-log scale. Red lines
represent fitted results.} \label{fig1}
\end{figure}

The fluctuation property $\sigma_r(k_0)$ can be calculated by
\begin{equation}
\sigma_r(k_0)\equiv \sqrt{<r^2(\Delta t)>-<r(\Delta t)>^2},
\end{equation}
where $<>$ represents the average taken for the same $k_0$ and the
observed period $\Delta t$ can be one year, one month or one day in
the present paper. In Fig.~\ref{fig1}(a), we plot the conditional
mean of $r$ for different periods. All of them are around constant
zero, which is independent of the initial degree. However the
conditional deviation of the three periods display different
behaviors, as is shown in Fig.~\ref{fig1}(b) $\thicksim$
Fig.~\ref{fig1}(d). For daily fluctuation, $\sigma_r(k_0)$ decays as
power law with exponent about $-0.5$, which coincides with the
prediction of PA rule. For monthly fluctuation, the small-degree
region of $k<10$ becomes flat while the rest of region remains
unchanged. With the increase of $\Delta t$, the flat area extends
gradually and Gibrat law becomes dominated for a large region of
$k<300$ for yearly fluctuation. These results indicate a crossover
transition from PA to Gibrat phase, which clearly rejects the
stationarity of the fluctuation. Therefore neither PA nor Gibrat law
can characterize the overall fluctuation property of Internet. They
validate only for a specific period. For short period PA matches
while for long period Gibrat law takes over. Note that our finding
is different from those of human dynamics and firm growth, where a
single universal scaling law is reported for the whole conditional
deviation~\cite{human,stane1,stane3}.

\begin{figure}[htbp]
\centering
\includegraphics[width=2.5in]{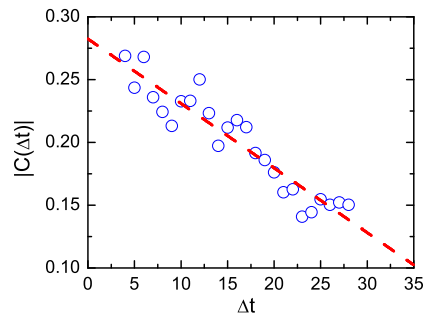}
\caption{Absolute value of the average correlation between $k_0 $
and $\sigma_r(k_0 )$ versus observed period $\Delta t$. It follows
approximately a linear decrease, which is fitted by the red dashed
line as $\mid C(\Delta t)\mid\approx -0.005\times \Delta t+0.28$.}
\label{fig2}
\end{figure}

To better understand the scope of the application of the two
classical models, we need to determine their validated periods
$\Delta t_v$.  For PA, we find the corresponding $\Delta t_v$ is no
more than several-day magnitude and we can affirm that PA is always
valid for $\Delta t_v<1$day as is indicated in Fig.~\ref{fig1}(b).
For Gibrat law, $\Delta t_v$ corresponds to when the correlation
coefficient of $\sigma_r(k_0)$ and $k_0$ is zero. Therefore we study
the relation between the correlation coefficient and the observed
period $\Delta t$ by using monthly data. Specifically, for a
particular $\Delta t$ we calculate the correlation coefficients
$C_{\sigma,k}(t,\Delta t)$ for all $t\in\{1,2,3,..,36-\Delta t)\}$
and average them so that
\begin{equation}
C(\Delta t)= \frac{\sum_{t=1}^{36-\Delta t}C_{\sigma,k}(t,\Delta
t)}{36-\Delta t}.
\end{equation}
We consider $C(\Delta t)$ characterize the general correlation
coefficient of $\sigma_r(k_0)$ and $k_0$ for the observed interval
$\Delta t$. We calculate Eq(12) for $\Delta t\in\{1,2,...,28\}$ and
present its absolute value in Fig.~\ref{fig2}~\cite{buchong}. We
find that despite large deviation for the results of $\Delta
t<4$(not plotted), the main body of $\mid C(\Delta t\geq4)\mid$
displays a linear decay which is fitted as
\begin{equation}
\mid C(\Delta t)\mid\approx -0.005\times \Delta t+0.28
\end{equation}
Then let $\mid C(\Delta t)\mid=0$, we can evaluate $\Delta t_v=
56$months$\approx4.6$years. Therefore the Gibrat law is expected to
be totally valid for at least $4.6$-year period. Note that $\Delta
t_v$ of Gibrat law estimated by using yearly data gives the similar
result even though the quality of the fitting is poorer due to much
smaller length of both $t$ and $\Delta t$.

\begin{figure}[htbp]
\centering
\includegraphics[width=2.5in]{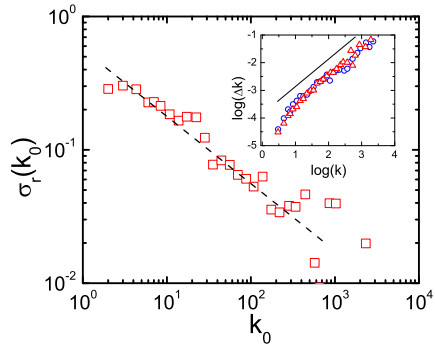}
\caption{The conditional standard deviation for the reshuffled
yearly data. It follows a power-law decay of exponent about $-0.5$,
as is indicated by the black dashed line. The result contrasts to
that of the original yearly data but is consistent with that of PA
and daily data. Inset: The empirical result of the proportionate
effect before(blue circle) and after(red triangle) the reshuffling
operation. The statistical analysis is based on mean-field treatment
as was done in previous studies~\cite{confirm4,confirm5,confirm6}.
The black solid line is of slope $1$, which is a guide for eyes.}
\label{fig3}
\end{figure}

\begin{figure}[htbp]
\centering
\includegraphics[width=2.65in]{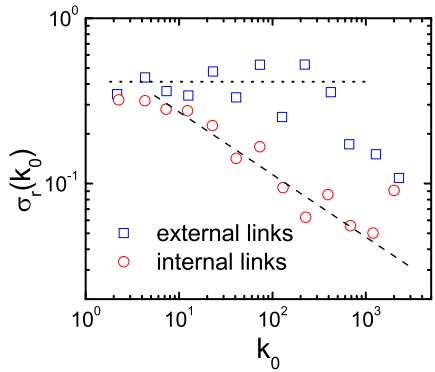}
\caption{The conditional standard deviation for reshuffling the
creation of external(blue square) and internal(red circle) links.
$\sigma_r(k_0)$ after reshuffling the external links appears no
significant difference from that of the original yearly data.
However $\sigma_r(k_0)$ after reshuffling the internal links decays
with power-law exponent about $-0.43$. The dotted line is horizontal
while the dashed one is the fit line.} \label{fig4}
\end{figure}

The crossover transition indicates that there are some underlying
mechanism that give rise to the emergence of Gibrat law. Reminding
that memoryless and independent creation of links can only cause a
power-law decay with an exponent of $-0.5$ of $\sigma_r(k_0)$,
Gibrat law with constant conditional standard deviation probably
indicates the existence of strong correlation in the evolution of
Internet. Indeed studies on population growth and human
communication dynamics demonstrated that correlation could lower the
related power-law exponent~\cite{stane2,human,human2}. This
speculation can be confirmed by reshuffling the creation of links
for yearly data. In specific, we change randomly the order of the
creation of links while maintain the topology of first year($1998$)
and last year($2012$). We first check whether the reshuffling
operation changes the basic evolution mechanism, i.e. the
proportionate effect. Surprisingly, the proportionate effect $\Delta
k\sim k$ maintains as before(inset of Fig.~\ref{fig3}), but the
fluctuation pattern of the degree growth rate changes from a
constant value to a power-law decay of exponent $-0.5$, which is
exactly the behavior of PA and daily data(Fig.~\ref{fig3}). This is
a direct evidence for the existence of the correlation and its
contribution to Gibrat law. Indeed the reshuffling process does not
change PA at mean-field level at all but only destroys any possible
correlation between the creation of links. Therefore we draw the
conclusion that correlation is the essential ingredient responsible
for the emergence of Gibrat law. The crossover transition thus
indicates that such a correlation occurs first at small-degree nodes
and spreads to large-degree nodes gradually. To further identify the
origin of Gibrat law, we separate the external links(links created
between new-coming node and old existing node) from internal
links(links created between existing old nodes) and reshuffle one
while maintain the other. As shown in Fig.~\ref{fig4}, reshuffling
the external links has little effect on the fluctuation pattern. On
the other hand reshuffling the internal links causes a clear power
law decay of the conditional standard deviation with exponents about
$-0.43$. Therefore we conclude that the major part of the
correlation comes from the internal links, which has more
contribution to the emergence of Gibrat law.

\section{Conclusion}
We have shown the inconsistency between PA and Gibrat law and
determine their scope of application to Internet. By analyzing the
conditional standard deviation of the degree growth rate, we find
that the actual fluctuation of Internet exhibits a crossover
transition from PA to Gibrat law with the increase of the observed
period. We have determined that the scope of the validation is about
several-day magnitude of period for PA while $4.6$-year of period
for Gibrat law. We briefly study the origin of the emergence of
Gibrat law and find it most related to the correlation between the
internal links.

There has been an argument that whether the construction of Internet
is governed by the randomness of self-organized nature or highly
designed order of engineered nature~\cite{internet0}. Although
self-organized system does not rule out the possibility of
correlation, the strong correlation found in the evolution of
Internet consists with the designed order of the engineered
intuition. The present empirical results indicate that purely random
description based on mean-field approach, which ignores correlation,
might match short-term(daily) Internet fluctuation, but is very
insufficient to characterize the long-term(yearly) evolution. The
crossover paradigm of the dynamical fluctuation provides a test that
any future model should pass. Therefore the consideration of memory
effect~\cite{shehui} as well as how such effect works is critical
for a complete Internet model theory.

This work was partially supported by the National Nature Foundation
of China under Grant Nos. 11075057, 11035009 and 10979074.

 {}
\end{CJK*}

\begin{thebibliography}{}
\bibitem{buchong} As expected the correlation coefficients are all negative. Taking
the absolute values does not change its trends at all. we choose
$\Delta t<28$ because larger $\Delta t$ causes the length of $t$ too
small to apply statistical analysis.
\end{thebibliography}
\end{document}